\documentclass[11pt]{article}

\usepackage[iso-8859-7]{inputenc}
\usepackage{epsfig}
\usepackage{graphicx}

\usepackage{amsfonts}
\usepackage{amssymb}
\usepackage{amsthm}
\usepackage{amsmath}
\usepackage{multirow}
\usepackage{cite}

\usepackage{mathtools}% ENUMERATES THE LABELED EQUATIONS THAT ARE REFFERED TO
\newcommand{\newc}{\newcommand}
\newc{\ra}{\rightarrow}
\newc{\lra}{\leftrightarrow}
\newc{\ov}{\overline}
\newc{\pa}{\partial}

\newc{\be}{\begin{equation}}
\newc{\ee}{\end{equation}}

\newc{\ba}{\begin{eqnarray}}
\newc{\ea}{\end{eqnarray}}

\newc{\n}{\nu}
\newc{\D}{\Delta}
\newc{\eps}{\epsilon}
\newc{\la}{\lambda}
\newc{\e}{\epsilon}
\newc{\nn}{\nonumber}
\newcommand{\ga}{\alpha}

\begin{document}
%\mathtoolsset{showonlyrefs=true}%ENABLES THE mathtool
\thispagestyle{empty}

\vskip 2truecm
%%%%%%%%%%%%%%%%%%%%%%%%%%%%%%%%%%%%%%%%%%%%%%%%%%%%%%%%%%%%%%%%%%%%%%%%%%%%%%%%%%%%%%%%%%%%%%%
\vspace*{3cm}
\begin{center}
{\large{\bf
Unification, KK-thresholds and the top Yukawa coupling  in {\cal F}-theory  GUTs}}

\vspace*{1cm}
{\bf G.K.  Leontaris$^{(1)}$, N.D. Tracas$^{(2)}$ and G. Tsamis$^{(2)}$}\\
$^1$Theoretical Physics Division, Ioannina University, GR-45110 Ioannina, Greece\\
$^2$Physics Department, National Technical University, 157 73 Athens, Greece
\end{center}

\vspace*{1cm}
\begin{center}
{\bf Abstract}
\end{center}

\noindent
In a class of F-theory $SU(5)$ GUTs the low energy chiral mass spectrum is obtained
from rank one fermion mass textures with a hierarchical structure organised by $U(1)$
symmetries embedded in the exceptional $E_8$ group.  In these theories chiral fields
reside on matter `curves'  and the tree level masses are computed from integrals of
overlapping wavefuctions of the particles at the triple intersection points. This
calculation requires knowledge of the exact form of the wavefuctions.
In this work we propose a way to obtain a reliable estimate of the various quantities
which determine the strength of the Yukawa couplings.  We use previous analysis of
KK threshold effects to determine the (ratios of)
heavy mass scales of the theory  which are involved in the
normalization of the wavefunctions.  We consider similar effects from the chiral spectrum
of these models and discuss possible constraints on the emerging matter content.
In this approach, we find that the Yukawa couplings can be determined solely from the $U(1)$ charges of the
 states  in the `intersection' and  the torsion which is a topological  invariant quantity.
We apply the results to a viable $SU(5)$ model with minimal spectrum which satisfies
all the constraints imposed by our analysis. We use renormalization group analysis to
estimate the  top and bottom masses and find that they are in agreement with the
experimental values.

\vfill
\newpage
\section{Introduction}

Recent progress in F-theory model
building~\cite{Donagi:2008ca,Beasley:2008dc,Beasley:2008kw,Hayashi:2008ba,Donagi:2008kj,Heckman:2009mn,Marsano:2009wr,
Blumenhagen:2009up}\footnote{
For other recent related work and a review see also~\cite{Grimm:2010ks,Weigand:2010wm}}
has shown that old successful GUTs, including
the Georgi-Glashow minimal SU(5),  the SO(10) model etc, are naturally realised on the world-volume of non-perturbative
seven branes wrapping appropriate compact surfaces. The rather interesting fact in F-theory constructions is that
they are defined on a compact elliptically fibered Calabi-Yau complex four dimensional manifold thus the exceptional groups
$E_6,E_7, E_8$, can be naturally incorporated into the theory too~\cite{Donagi:2008ca,Beasley:2008dc,Beasley:2008kw,Heckman:2009mn}.
Although exceptional gauge symmetries suffer from several drawbacks when realized in the context of four-dimensional
grand unified theories, in the case of F-theory models they are more promising as new possibilities arise for the
symmetry breaking mechanisms and the derivation of the desired massless spectrum.

Present studies on F-theory model building have been concentrated  on three generation -mainly $SU(5)$- GUT models
which fall into the following two distinct categories:
those where all three families with the same Standard Model representation content are assigned
to a single matter curve~\cite{Heckman:2008qa,Cecotti:2009zf,Marsano:2009gv,Conlon:2009qq},
and variants~\cite{Dudas:2009hu,King:2010mq,Dudas:2010zb,Leontaris:2010zd,Hayashi:2009bt,Grimm:2010ez,Ludeling:2011en}
where some or all of the quark and lepton families are assigned to different curves. Several of these constructions
built up to these days have attempted to give solutions to fundamental GUT problems as is the case of doublet-triplet
splitting, the rapid proton decay, the Higgs mixing term, the neutrino sector and other related
 issues~\cite{Beasley:2008kw,Heckman:2009mn,King:2010mq,Weigand:2010wm,Ludeling:2011en,Li:2010zz,Kuflik:2010dg}.
To analyse the phenomenological properties one should extract the relevant information from the superpotential
which can be readily constructed once a particular assignment of the fermion families and Higgs on the  matter
curves has been chosen. Of course,  dominant r\^ole on the estimation of such effects is played by the Yukawa
couplings, thus the theory's predictive power depends on the calculability of the latter.

In F-theory GUTs  the trilinear Yukawa couplings are realised at the intersections of three matter curves
$\Sigma_i,\;{i=1,2,3}$ where the zero-modes of two fermion fields and a Higgs boson reside. Along these curves
the $G_S$ symmetry is enhanced $G_{\Sigma_i}\supset G_S\times U(1)_i$ while the corresponding  zero modes are
charged under the $U(1)_i$. To determine the most general structure of the zero-mode wavefunctions one has
to solve their corresponding differential equations of motion emerging from the twisted eight-dimensional Yang-Mills
action~\cite{Beasley:2008dc},  (see also \cite{Hayashi:2009ge,Conlon:2009qq,Font:2009gq, Marchesano:2010bs}).
In general, the solutions  are found
to exhibit the expected gaussian~\cite{Hayashi:2009ge,Font:2009gq}
profile which falls off exponentially away from the curve while their exact form is specified by a mass scale
characterizing the size of the compact space and the particular $U(1)_i$-charge of the relevant zero-mode.
The Yukawa couplings of the $\{33\}$-entries of the up, down and charged lepton mass matrices are then
computed in terms of the integrals of overlapping wavefunctions of the aforementioned  form at the intersection
point of three matter curves
\be
\label{yukawa}
\lambda_{ij}\propto\,M_*^4\int_S\psi_i\psi_j\phi\, dz_1\wedge\,d\bar z_1\wedge dz_2\wedge\,d\bar z_2
\ee
where the integration is over the compact internal four-dimensional manifold $S$
supporting the gauge group of the theory,
$\psi,\phi$ denote the fermion and Higgs wavefunctions respectively whilst $M_*$ is associated to
the supergravity limit of the compactified theory, hence it will be naturally linked to the
F-theory compactification scale. Furthermore, assuming that higher order non-renormalizable Yukawa
couplings are generated through mediation of heavy string modes and/or Kaluza-Klein states, the
calculation of  all Yukawa entries of this  type  in the fermion mass matrices can be reduced to a
similar computation~\cite{Leontaris:2010zd}.

To quantify low energy implications of a given model, a reliable estimation of the integral (\ref{yukawa})
is required. This is however a non-trivial task since, although the local profile of the wavefunction
is precisely known, the final result is expressed in terms of several parameters which mainly appear
in the normalization of the wavefunction. In particular, the background scalar field vev, the  $M_{GUT}$ scale, the scale
$M_*$ as well as the  geometry of the  surface $S$ supporting the GUT
symmetry are not precisely known. Estimates of such global quantities have been given and  rather sensible results have
been obtained in the case of the mass hierarchy~\cite{Cecotti:2009zf,Dudas:2009hu,Leontaris:2010zd}.
However,  in this context one fails to predict a heavy top mass compatible with the experimental data in an unambiguous manner.

In this work we reconsider the problem of trilinear Yukawa couplings in F-theory GUTs, focussing on the
normalization of the wavefunction and the r\^ole of the various scales of the theory.  We rely on
the results of~\cite{Friedmann:2002ty} and~\cite{Donagi:2008kj} to determine the effects of
KK-massive modes on the renormalization group running the GUT scale and the gauge unification.
Using these  results we estimate the modifications of  mass scales ratio  $M_{GUT}/M_C$
and express this solely in terms of the Ray-Singer torsion~\cite{Ray:1973sb} which is a topologically invariant
quantity.  Identifying the various scales appearing  in the computation of the overlapping integrals with
$M_{GUT}$ and $M_C$, we argue that this gives a reliable computation for the top Yukawa coupling.
The final result for the Yukawa coupling depends only on the charges of the fields involved and
an exponential factor involving the torsion.

In the next section we review in brief the basic F-theory set up and the relevance of the Ray-Singer
torsion to the threshold corrections. We show how KK-massive modes modify the GUT scale
and in section 3 we perform a similar analysis for the case of chiral matter subject to flux
constraints.  In section 4 we apply the analysis in a realistic F-theory $SU(5)$ GUT~\cite{Leontaris:2010zd}
which was shown to reproduce successfully the fermion hierarchy and confront other main phenomenological
issues. In sections 5 and 6 we perform the renormalization group analysis
and obtain the formulae for the top and bottom Yukawa couplings by computing the corresponding integrals.
As already said the result is expressed in terms of the analytic torsion. As an application,
we perform an explicit numerical calculation of  the  top and bottom Yukawa  couplings choosing
a specific  line bundle on a Hirzebruch surface. In section 7 we summarize our work and present our conclusion.
Some technical details regarding the solutions of the
renormalization group equations for the computation of the third generation quark masses are shown in the appendix.

\newpage
\section{KK-modes and the GUT scale}

Our F-theory set up is assumed to be the world-volume of the seven-brane  of
a ADE-type singularity which wraps the space $R^{3,1}\times S$ where  $S$  is
a K\"ahler manifold of two complex dimensions.  At low energies,
F-theory is described by an eight-dimensional  Yang-Mills theory on $R^{3,1}\times S$
which must be topologically twisted to preserve $N=1$
supersymmetry~\cite{Beasley:2008dc,Beasley:2008kw}. The supersymmetric multiplet consists
of the gauge field, two scalars $\Phi_{8,9}$ combined into the complex fields $\phi/\bar\phi=\Phi_8\pm i\Phi_9$
and the fermions $\eta,\psi,\chi$ all in the adjoint of the gauge group.
In the twisted theory  $\eta,\psi,\chi$  appear as holomorphic $(k,0)$ forms (and their complex conjugate
antiholomorphic $(0,l)$)  with $k,l=0,1,2$, while the scalar $\phi$ is a two form. At the $d=4$, $N=1$ level,
these fields are organised as one gauge  and two chiral multiplets
\[
(A_{\mu},\eta), (A_{\bar m},\psi_{\bar m}), (\phi_{12},\chi_{12})
\]

 For the computation of threshold corrections we assume that the  gauge theory model is
described by some GUT gauge group which for definiteness we take it to be  $G_S=SU(5)$
(at a scale  $M_{GUT}\ll M_{Planck}$).   In this limit it is also
natural to assume that KK-modes are  much lighter than other string excitations.

Threshold corrections are expressed in terms of the  masses of the KK-spectrum which emerge
as the  non-zero eigenvalues of the Laplacian of the eight dimensional theory. Therefore,
threshold corrections are constants  and should be independent of the particular
metric on the surface $S$, while they are expected to be expressed in terms of a
topologically invariant quantity. It was first shown, already
sometime ago~\cite{Friedmann:2002ty}, that in the context of $M$-theory compactified
on a manifold with $G_2$ holonomy the most appropriate topological
invariant quantity to express these threshold corrections is the Ray-Singer
torsion~\cite{Ray:1973sb}.  A similar analysis has been performed for the case of
$F$-theory in~\cite{Donagi:2008kj} which we now review in brief.

The $SU(5)$ subgroup which commutes with the
Standard Model ($G_{SM}$) is the $U(1)_Y$.  We denote with
$q_i$ the  $U(1)_Y$ charge  of the Standard Model
representations $R_i$ arising from the decomposition of the adjoint
of $SU(5)$ under $G_{SM}$
\begin{align}
\label{AdjSU5}
\mathbf{24\rightarrow (8,1)_0+(1,3)_0+(1,1)_0+(3,2)_{-5/6}+(\bar 3,2)_{5/6}}
\end{align}
We assume that the GUT group breaks to $G_{SM}$ by turning on a flux along $U(1)_Y$,
hence each $q_i$ determines now a line bundle which  we denote it by  ${\cal L}^{q_i}$.
We observe that in the $SU(5)$ adjoint decomposition (\ref{AdjSU5}), in addition to the
standard $SU(3),SU(2)$ adjoints there are in principle exotic zero modes in the $(3,2)$
and $(\bar 3,2)$ representations along the ${\cal L}^{\pm 5/6}$ line bundles
whose numbers are given by the Euler character $-\chi(S,{\cal L}^{\pm 5/6})$.
Therefore, elimination of these massless modes  requires $\chi(S,{\cal L}^{\pm 5/6})=0$
which implies~\footnote{The number of massless modes  of a ${\cal T}_i$ representation
is  $-\chi(S,{\cal T}_i) =1+\frac 12 c_1({\cal T}_i)\cdot (c_1({\cal T}_i)+c_1(S))$ while
here  the following relation is satisfied: $c_1({\cal L}^{5/6})\cdot c_1(S)=0$~\cite{Beasley:2008kw}.}
the following relation for the cohomology class: $c_1({\cal L}^{5/6})\cdot c_1({\cal L}^{5/6})=-2$.

Consider next the Dolbault operator $\bar\partial$ of the corresponding holomorphic bundle $V$ with
representation  $ R(V)$, acting on $k$-forms as follows
\be
\label{Dolbault}
\bar\partial :\Omega_S^{0,k}\otimes R(V)\ra\Omega_S^{0,k+1}\otimes R(V),\;{\rm for}\;k=0,1
\ee
and  the Laplacian
\be
\label{Laplacian}
\Delta_{k,R(V)}=(\bar\partial+\bar\partial^{\dagger})^2=\bar\partial\bar\partial^{\dagger}+
\bar\partial^{\dagger}\bar\partial
\ee
If we denote collectively with $\psi_k^{n}$ its $k$-form eigenfunction then
\be
\label{Eig}
\Delta_{k,R(V)}\psi_k^{n}=\lambda_n^k\psi_k^{n}
\ee
where $\lambda_n^k$ represents the corresponding eigenvalue and in four dimensions
corresponds to a squared mass.

We express now the threshold corrections of the gauge multiplet in terms of the Ray-Singer torsion.
The running of the gauge couplings are governed by the equation
\be
\label{gauge}
\frac{16\pi^2}{g^2_a(\mu)}=\frac{16\pi^2
k_a}{g^2_s}+b_a\log\frac{\Lambda^2}{\mu^2}+\mathcal{S}_a^{(g)},\quad
a=3,2,Y
\ee
where $\Lambda$ is the gauge theory cutoff scale, $k_a=(1,1,5/3)$ are the
normalization coefficients for the usual embedding of the Standard
Model to $SU(5)$, $g_s$ is the gauge coupling as deduced from the
higher theory and $b_a$ are the one-loop $\beta$-function
coefficients  which are given
by~\footnote{We adopt here the notation of ~\cite{Friedmann:2002ty}.}
\be
\label{beta}
b_a=2\,\textrm{Str}_{M=0}Q^2_a\left(\frac{1}{12}-\chi^2\right)
\ee
where $\chi$ is the helicity operator, $\textrm{Str}$ denotes the
supertrace (bosons contribute with weight $+1$ and fermions with
$-1$) and $Q_a$ stands for the three generators of the
Standard Model gauge group $SU(3)\times SU(2)\times U(1)_Y$.
Finally, $\mathcal{S}_a^{(g)}$ correspond to the one-loop threshold
corrections and are given by a similar term over the massive states
(in contrast with the previous term that corresponds to massless
states)
\begin{align}
S_a^{(g)}&=2\,\sum_{R_i} \textrm{Str}_{M\neq 0}Q^2_a\left(\frac{1}{12}-\chi^2\right)\log(\Lambda^2/M^2)\label{KK-contr1}\\
&=2\,\sum_{i}\textrm{Tr}_{R_i}Q^2_a \,\,\,\textrm{Str}_{M\neq 0}\left(\frac{1}{12}-\chi^2\right)\log(\Lambda^2/M^2)\label{KK-contr2}
\end{align}
In the second line $\textrm{Str}_{M\neq 0}Q^2_a$ has been factored out since it depends only
on the representation $R_i$ (we are suppressing the notation for the bundle $V$ in $R_i(V)$).
For a certain helicity state there is a logarithmic dependence on its mass squared $M^2$ which
corresponds to the eigenvalue of the Laplacian  $\Delta_{k,R_i}$. Further, since the trace of $\log$
is equal to the logarithm of the determinant we can replace the trace of $\log\Lambda^2/M^2$ above with
\be
\label{Delta}
-\log\det{'}(\Delta_{k,R_i}/\Lambda^2)
\ee
where the prime on $\det$ denotes that we exclude  the zero modes.

Now, each eigenvector of the zero-form Laplacian $\Delta_{0,R_i}$ contributes a vector multiplet with
helicities $1,-1,\frac 12,-\frac 12$, while
the one-form Laplacian $\Delta_{1,R_i}$ gives a chiral multiplet with helicities $0,0,\frac 12,-\frac 12$.
Similarly,  $\Delta_{2,R_i}$ is associated to anti-chiral multiplets.
Evaluating the supertrace we get
\begin{align}
\textrm{Str}\left(\frac{1}{12}-\chi^2\right)&
                                         =\left\{
                                         \begin{array}{ll}
                                          -\frac32&\textrm{for the vectot multiplet}\\
                                          \\
                                           + \frac 12&\textrm{for the chiral multiplet}
                                           \end{array}
                                           \right.
                                            \label{chi-c}
\end{align}
Therefore, the total sum of the contribution of KK-modes to the thresholds \textit{from the gauge fields}
is then written
\be
\label{KK-contr3}
\mathcal{S}_a^{(g)}=2\sum_{i}{\rm Tr}_{R_i} (Q_a^2){\cal K}_i
\ee
with
\be
\label{KK-contr4}
{\cal K}_i=\frac{3}{2}\log \det{'}\frac{\Delta_{0,R_i}}{\Lambda^2}-\frac 12\log \det{'} \frac{\Delta_{1,R_i}}{\Lambda^2}
-\frac 12\log \det{'} \frac{\Delta_{2,R_i}}{\Lambda^2}
\ee
and
\ba
\textrm{Tr}_{R_0} \left(Q^2_{\left\{3,2,Y\right\}}\right)&=
\left\{\begin{array}{ll}\{3,2,0\},& \textrm{for}\quad
\mathbf{(8,1)_0+(1,3)_0+(1,1)_0}\\
\left\{1,\frac 32,\frac{25}6\right\},& \textrm{for}\quad
\mathbf{(3,2)_{-5/6}\,,\,(\bar 3,2)_{5/6}}\end{array}
\right.
\label{Q22}
\ea
with normalization ${\rm Tr}(Q_a^2)=\frac{k_a}2$. Therefore, the three $\mathcal{S}_a^{(g)}$ read
\be
\label{S}
\begin{split}
\mathcal{S}^{(g)}_Y&=\frac{25}{3}\left({\cal K}_{5/6}+{\cal K}_{-5/6}\right)\\
\mathcal{S}^{(g)}_2&=4{\cal K}_0+3\left(K_{5/6}+{\cal K}_{-5/6}\right)\\
\mathcal{S}^{(g)}_3&=6{\cal K}_0+2\left({\cal K}_{5/6}+{\cal K}_{-5/6}\right)
\end{split}
\ee
with an obvious notation in the subscript of ${\cal K}$.
Further, the spectrum of $\Delta_1$ is equivalent to the sum of the spectra of $\Delta_0$ and $\Delta_2$.
Then we can write (\ref{KK-contr4}) as follows
\be
\label{KK-contr4A}
{\cal K}_i={2}\log \det{'}\frac{\Delta_{0,R_i}}{\Lambda^2}-\log \det{'} \frac{\Delta_{1,R_i}}{\Lambda^2}
\ee
Introducing the Ray-Singer torsion
\be
\label{torsion}
{\cal T}_R=\frac 12\sum_{k=0}^2(-1)^{k+1}\log \det{'}\frac{ {\Delta_{k,R_i}}}{\Lambda^2}
\ee
and noting that the Laplacian commutes with the Hodge $*$ operator which maps
$k$-forms to $(2-k)$-forms, so that $\Delta_{k,R}$ and $\Delta_{2-k,R^*}$ have the same spectrum,
we get~\cite{Donagi:2008kj}
\begin{equation}
\label{torsion2}
{\cal T}_R=-{\cal T}_{R^*}=\frac{1}{2}\left(2 \log \det{'}\frac{{\Delta_{0,R_i}}}{\Lambda^2}-\log \det{'} \frac{{\Delta_{1,R_i}}}{\Lambda^2}\right)
\end{equation}
The expression inside the parenthesis is exactly what we have derived previously for the
threshold corrections.
According to the Ray and Singer theorem~\cite{Ray:1973sb}  the torsion is independent
of the metric of the manifold and therefore independent of the cutoff scale $\Lambda$. (The
latter can be eliminated by scaling the metric.)

In the above expression for the torsion it has been assumed that there are no zero modes as is
the case of a non-trivial  representation $R_i$. However, in the case of the trivial bundle
there are zero modes and the torsion is not equal to ${\cal K}_{R_0}$. These are related by
\begin{equation}
\label{zero_modes}
2{\cal T}_{\cal O}={\cal K}_{\cal O}-\log(V\,\Lambda^2)
\end{equation}
where in the subscript we used the standard notation ${\cal O}$ for the trivial representation.

We assume  $S$ to be a del Pezzo surface and we expect that ${\cal K}_{5/6}={\cal K}_{-5/6}$ so that
the equations in (\ref{S}) can be cast in the form
\begin{equation}
\label{S56}
\left({\cal S}_Y^{(g)},{\cal S}_2^{(g)},{\cal S}_3^{(g)}\right)=
\left(\frac{50}{3}{\cal K}_{5/6},\,6{\cal K}_{5/6}+4{\cal K}_0,\,4{\cal K}_{5/6}+6{\cal K}_0\right)
\end{equation}
Taking into account that the contribution of the gauge multiplet to the $\beta$-function
is $b_a^{(g)}=(0,-6,-9)$ we can rewrite the above as
\begin{equation}
\label{S56_2}
{\cal S}^{(g)}_a=\frac 23 b_a^{(g)} \left({\cal K}_{5/6}-{\cal K}_0\right)+10k_a{\cal K}_{5/6}=
\frac 43 b_a^{(g)} \left({\cal T}_{5/6}-{\cal T}_0\right)+20\,k_a{\cal T}_{5/6}
\end{equation}
The second term is  proportional to $k_a$ and therefore it can be  absorbed to the $16\pi^2k_a/g^2_s$ term,
redefining the coupling ${g}_s$. And, finally, the initial equation (\ref{gauge}) for the running of the gauge
coupling, where only the gauge multiplet has been taken into account, gives
\begin{equation}
\label{gauge2}
\begin{split}
\frac{16\pi^2}{g^2_a(\mu)}&=\frac{16\pi^2 k_a}{g^2_s}+b_a^{(g)}\log\frac{\Lambda^2}{\mu^2}+\mathcal{S}_a^{(g)}\\
&=\frac{16\pi^2 k_a}{g^2_s}+
b_a^{(g)}\log\frac{\exp\left[4/3\left({\cal T}_{5/6}-{\cal T}_0 \right)\right]}{\mu^2 V^{1/2}}
\end{split}
\end{equation}
where all the $\Lambda$ dependence, after the appropriate cancellation between the massless and massive modes,
has been absorbed in the redefinition of the bare gauge coupling along with the second term of (\ref{S56_2}),  and we can define the $M_{GUT}$ as
\begin{equation}
\label{MGUT}
M_{GUT}^2=\frac{\exp\left[4/3\left({\cal T}_{5/6}-{\cal T}_0 \right)\right]}{V^{1/2}}
\end{equation}
If we associate the world volume factor $V^{-1/4}$ with the characteristic
compactification scale $M_C$, we can write this equation as follows
\begin{equation}
\label{MGUT1}
M_{GUT}=e^{2/3\left({\cal T}_{5/6}-{\cal T}_0 \right)}\,M_C\,\cdot
\end{equation}
Thereby, we conclude that the ratio $M_{GUT}/M_C$ of the two characteristic scales of the theory depends
only on the torsion and is topologically invariant.

\newpage

\section{The inclusion of chiral matter}

We have seen in the previous section that the contribution of threshold corrections from
KK-modes of the vector multiplet leads to a cutoff independent
RG running of the gauge couplings. We wish now to include the analogous contributions from
the chiral and Higgs sector of the theory. In F-theory constructions we need to take into account
the zero mode as well as the KK-massive mode contributions.

We first start  by incorporating the zero mode effects.
In the case of the $SU(5)$ theory we discuss in this paper the possible non-trivial representations transform as $10,\overline{10}$ and $5,\bar 5$.
Matter fields in general, will contribute to the RG running in (\ref{gauge2}) with a term  of the form $b_a^x\log\frac{{\Lambda'}^2}{\mu^2}$ where
$b_a^x$ are the $\beta$-function coefficients for the corresponding  three SM gauge group factors and $\Lambda'$ a cutoff scale not necessarily
equal to the gauge cutoff $\Lambda$.  If matter fields in the theory arise in complete $SU(5)$ multiplets, then the $\beta$-functions
contribute in proportion to the coefficients $k_a$, ie. $b_a^x\propto k_a$.
Then  writing $\log\frac{{\Lambda'}^2}{\mu^2}$ as $\log\frac{{\Lambda'}^2}{M_{GUT}^2}+
\log\frac{M_{GUT}^2}{\mu^2}$ we can absorb  the first logarithm in a redefinition of the gauge coupling
\begin{equation}
\label{redefaG}
\frac{16\pi^2}{g_{GUT}^2}\,k_a\equiv \frac{4\pi}{a_{GUT}}\,k_a=\frac{4\pi}{a_S}\,k_a+b_a^x\log\frac{\Lambda'^2}{M^2_{GUT}}
%\frac{4\pi}{g_{GUT}^2}&\equiv\frac{1}{a_{GUT}}=\frac{1}{a_s}+\frac{b_a^x}{4\pi k_a}\log\frac{{\Lambda'}^2}{M_{GUT}^2}
\end{equation}
The massless chiral and Higgs spectrum at low energies, however, does not form complete
$SU(5)$ multiplets. As it is well known, Higgs doublets are found in $5+\bar 5$ representations
of $SU(5)$ together with the dangerous color triplets. To protect proton from rapid decay, triplets
must receive a mass at a relatively high scale $M_X\le M_{GUT}$. Taking this into account
we write $b_a^x=b_a+b_a^T$ where $b_a$ denotes the MSSM $\beta$-functions  while $b_a^T$
the color triplet part. Adding all contributions, (\ref{gauge2}) takes the form
\begin{equation}
\label{gauge3}
\begin{split}
\frac{16\pi^2}{g^2_a(\mu)}&=k_a\frac{16\pi^2}{g^2_{GUT}} +
(b_a^{(g)}+b_a)\log\frac{M_{GUT}^2}{\mu^2}+b^T_a\log\frac{M_{GUT}^2}{M_X^2}
\end{split}
\end{equation}
It should be mentioned that if the contributions from matter fields fail to obey the condition $b_a^x\propto k_a$,
then (\ref{redefaG}) splits the common gauge coupling\footnote{In
the next sections we  will see however that gauge coupling splitting at $M_{GUT}$ is not avoided when a
non-trivial $U(1)_Y$ flux is turned on to break  $SU(5)$.} to three
different values at $M_{GUT}$.

We discuss now the threshold contributions  arising from the $\Sigma_{\bar 5}$ and $\Sigma_{10}$ matter curves. From the decompositions
\[
{\bf 10\ra (3,2)_{\frac 16}+(\bar 3,1)_{-\frac 23}+(1,1)_1,\quad
 {\bar 5}\ra (\bar 3,1)_{\frac 13}+(1,2)_{-\frac 12}}
 \]
we readily find that KK-modes residing on these curves contribute to thresholds as follows~\cite{Donagi:2008kj}
\begin{equation}
\label{ga}
\begin{split}
S_a^{\bar 5}&=\left({\cal K}_{1/3},\;{\cal K}_{-1/2},\;{\cal K}_{-1/2}+2/3\,{\cal K}_{1/3}\right)\\
S_a^{10}&=\left(2{\cal K}_{1/6}+{\cal K}_{-2/3},\,3{\cal K}_{1/6},\;1/3\,{\cal K}_{1/6}+2{\cal K}_1+8/3{\cal K}_{-2/3}\right)
\end{split}
\end{equation}
where, as previously, ${\cal K}_i$ stand for the Ray-Singer torsion.  As in the case of gauge fields one
can  absorb the parts proportional  to $k_a$  by redefinition of the initial gauge coupling. Further, it is
anticipated that on appropriate bundle  structures such corrections will diminish or
be negligible compared to other (from light degrees of freedom or supersymmetry) threshold
effects~\cite{Blumenhagen:2008aw}-\cite{Jiang:2009za}.

In F-theory constructions one of the possible ways to break the GUT symmetries is to turn on a flux on the worldvolume of the seven-brane supporting
the unified gauge group.  In the present case, the $SU(5)$ gauge symmetry can be broken by turning on a non-trivial flux along the hypercharge
with $Q_Y={\rm diag}\{-\frac 13,-\frac 13,-\frac 13,\frac 12,\frac 12\}$.  As a result,  $SU(5)$ multiplets residing on certain curves where the flux
restricts non-trivially, might split. In the case of Higgs fiveplets in particular, this mechanism could be used to eliminate the unwanted triplets
from the light spectrum, however the effective theory may no longer contain complete $SU(5)$ representations and the  proportionality conditions
$b_a^x\propto k_a$ may not hold. We analyse this issue in the remaining of this section.

Suppose that $M_{10},M_5$  are two integers representing the number of complete SU(5)  $10$ and $\bar 5$ representations
in a specific construction. We work in the context of spectral cover approach, thus we consider the $E_8$ embedding   of $SU(5)_{GUT}$
and the breaking
\[
E_8\ra SU(5)_{GUT}\times U(1)^4
\]
The $SU(5)$ chiral and Higgs matter fields descend from the adjoint representation of the $E_8$ symmetry
and reside on the various curves denoted with $\Sigma_{10_j},\Sigma_{\bar 5_i}$.
We expect that the $U(1)$ fluxes (those not included in $SU(5)_{GUT}$) together with the
tracelessness condition $\sum_iF_{U(1)_i}=0$ imply the  following condition on the numbers of multiplets~\cite{Dudas:2010zb}
\begin{equation}
\label{trace0}
\sum_i M_5^i+\sum_j M_{10}^j=0
\end{equation}
Consider first the case that we have all $10$ type chiral matter  accommodated only on one $\Sigma_{10}$ curve
and all chiral states $\bar{5}$ respectively
on a single $\Sigma_{\bar 5}$ curve. Then condition (\ref{trace0}) implies the relation $M_{10}=-M_5=M$.

We denote with $N_{Y_5}, N_{Y_{10}}$ the corresponding units of $Y$ flux which splits the $SU(5)$ multiplets according to
\begin{equation}
\label{model_dif}
\begin{split}
\Sigma_{\bar 5}:\left\{
\begin{array}{l}
n_{(\mathbf{3,1})_{-1/3}}-n_{(\mathbf{\bar{3},1})_{1/3}}=M_5\\
n_{(\mathbf{1,2})_{1/2}}-n_{(\mathbf{1,2})_{-1/2}}=M_5+N_{Y_5}
\end{array}
\right.
\Sigma_{10}:\left\{
\begin{array}{l}
n_{(\mathbf{3,2})_{1/6}}-n_{(\mathbf{\bar{3},2})_{-1/6}}=M_{10}\\
n_{(\mathbf{\bar{3},1})_{-2/3}}-n_{(\mathbf{3,1})_{2/3}}=M_{10}-N_{Y_{10}}\\
n_{(\mathbf{1,1})_{1}}-n_{(\mathbf{1,1})_{-1}}\hspace*{.53cm}=M_{10}+N_{Y_{10}}
\end{array}
\right.
\end{split}
\end{equation}
Notice that these formulae count the number of $5$-components minus those of $\bar 5$
and the number of $10$ components minus those of $\overline{10}$. Since we know that
families are accommodated on $\bar 5$'s
 we require  $n_{(\mathbf{\bar{3},1})_{1/3}}>n_{(\mathbf{3,1})_{-1/3}}$ which implies
$M_5<0$. Similarly, because the remaining pieces of fermion generations live on $10$'s,
 we wish to end up with $10$-components after
the symmetry breaking, hence we should have $M_{10}>0$. For example, for exactly three generations
we should demand $M_{10}=-M_{5}=3$ and $N_{Y_j}=0$.  In general however, various curves  belong to
different homology classes and flux restricts non-trivially to some of them, thus $N_{Y_j}\ne 0$
at least for some values of $j$. Furthermore, we note that  in the above description
we deal with negative numbers which are used to discriminate  the $\bar{5},\overline{10}$
conjugate representations from $5$'s and  $10$'s.
In order to take correctly into account their contributions to the $\beta$-functions, care
should be taken so that only positive numbers are introduced. Thus for the number of triplets  we take $|M_5|=M$, while
for the required lepton doublets we take $|M_5+N_Y|=M-N_Y>0$ and so on~\footnote{
In fact here we assume that $|M_5|$ counts
exactly the number of  ${(\mathbf{\bar{3},1})_{1/3}}$ and that there are  no extra ${(\mathbf{3,1})_{-1/3}},{(\mathbf{\bar{3},1})_{1/3}}$ pairs
and similarly for the other components, otherwise threshold effects should be taken into account~\cite{Leontaris:2009wi,Palti:2009,Li:2010dp}.
More general cases are considered in the next sections.}.
 For this simple  example we observe that the contributions to $\beta$-functions come with the right proportionality
factors $1:1:\frac 53=k_3:k_2:k_1$ only in the unrealistic case $N_{Y_{5}}=N_{Y_{10}}=N$  where
$ \frac 35b_Y=b_3=b_2=2M-\frac 12\,N$,
so that  the arbitrary cutoff  $\Lambda'$  can be removed to a redefinition
of the gauge coupling as in the case of the gauge fields.

In the general case the situation is more complicated. In the $E_8$ embedded $SU(5)$ case in particular,
in principle there exist ten $\Sigma_{5}$ and five $\Sigma_{10}$
curves\footnote{This is clear from the decomposition of the adjoint of $E_8$, under $SU(5)\times SU(5)_{\bot}$
\[
248\ra (24,1)+(1,24)+(10,5)+(\overline{5},10)+(5,\overline{10})+(\overline{10},\overline{5})\nn
\].}
 where the corresponding
$SU(5)$ representations may reside. The implementation of monodromy actions may reduce the number
of these curves,  however, the construction of realistic effective GUT theories capable of reproducing the
know hierarchical  spectrum eventually require the involvement of  matter fields arising from a variety of matter curves.
Furthermore, because of $U(1)_Y$-flux effects several $SU(5)$ representations split in a
phenomenologically prescribed (and sometimes promiscuous) way resulting into incomplete multiplets.

Bearing in mind the above general context, we write the  contribution to the $\beta$-functions as
follows
\begin{equation}
\label{beta2}
\begin{split}
b_3&=\frac 12\sum_i|M_5^i|+ \sum_j|M_{10}^j|+\frac 12\sum_j |M_{10}^j-N_{10Y}^j|\\
b_2&=\frac 12\sum_i|M_5^i+N_{5Y}^i|+\frac 32\sum_j |M_{10}^j|\\
b_Y&=\frac 12\sum_i|M_5^i+N_{5Y}^i|+\frac 13\sum_i|M_5^i|\\
&\qquad\qquad+\frac 16\sum_j|M_{10}^j|+\frac 43\sum_j|M_{10}^j-N_{10Y}^j|+\sum_j|M_{10}^j+N_{10Y}^j|
\end{split}
\end{equation}
where $i,j$ summations are over the number of $\Sigma_{\bar 5}$ and $\Sigma_{10}$
discrete  curves in the quotient theory, i.e. after the monodromy action.

In order to acquire again the relation (1,1,5/3) we first must impose the equality between the contribution
to $SU(3)$ and $SU(2)$
\begin{equation}
\label{rel1}
\frac 12\sum_i|M_5^i|+ \sum_j|M_{10}^j|+\frac 12\sum_j |M_{10}^j-N_{10Y}^j|=
\frac 12\sum_i|M_5^i+N_{5Y}^i|+\frac 32\sum_j |M_{10}^j|\equiv {\cal B}
\end{equation}
where the parameter $ {\cal B}$ was introduced for later convenience. We may rewrite
the constraint above in a simplified way as follows
\begin{equation}
\label{rel2}
\sum_i|M_5^i|-\sum_i|M_5^i+N_{5Y}^i|=\sum_j|M_{10}^j|-\sum_j |M_{10}^j-N_{10Y}^j|
\end{equation}
Now, we can reexpress the $U(1)_Y$ contribution in terms of $ {\cal B}$ and a remaining
quantity  as follows
\begin{equation}\label{rel3}
\begin{split}
%&\frac 12\sum_i|M_5^i+N_{5Y}^i|+\frac 13\sum_i|M_5^i|+\\
%&\qquad\qquad\frac 16\sum_j|M_{10}^j|+\frac 43\sum_j|M_{10}^j-N_{10Y}^j|+\sum_j|M_{10}^j+N_{10Y}^j|=\\
b_Y&=\frac 53\,  {\cal B}-2\sum_j|M_{10}^j|+\sum_j|M_{10}^j+N_{10Y}^j|+\sum_j|M_{10}^j-N_{10Y}^j|
\end{split}
\end{equation}
In (\ref{rel3}) we have split the $U(1)$ $\beta$-function into two parts, the first being
$\frac 53{\cal B}$ which comparing with (\ref{rel1}) we observe that preserves the required
$k_a$-proportion  with $b_{2,3}$.
Therefore, the remaining part must be zero and finally we get  the following two constraints
\begin{equation}
\label{constraint}
\begin{split}
\sum_j|M_{10}^j|-\sum_i|M_5^i|&=\sum_j |M_{10}^j-N_{10Y}^j|-\sum_i|M_5^i+N_{5Y}^i|\\
2\sum_j|M_{10}^j|&=\sum_j|M_{10}^j+N_{10Y}^j|+\sum_j|M_{10}^j-N_{10Y}^j|
\end{split}
\end{equation}
In writing the above contributions to the $\beta$-functions we have assumed that they correspond to all the matter content,
including the 3 generations of the SM. Clearly, it is a non-trivial task to obtain a low energy
effective field theory model with only the MSSM spectrum.

We note in passing that if we want to consider only the extra matter content, we should subtract a 3 from all
absolute values: $|M_5^i|\rightarrow |M_5^i|-3$, $|M_5^i+N_{5Y}^i|\rightarrow |M_5^i+N_{5Y}^i|-3$, etc. But since we know
that the SM content forms a complete $\bar 5$ and a $10$, the contributions from the 3 generations obey the rule (1,1,5/3).
Therefore, the above constraints hold also for the case when only the extra matter is considered. Note also that if all $N_{Y_i}$
are zero the relations hold automatically, as they should.
Of course, any choice of $M$'s has to satisfy the constraint (\ref{trace0}).

\section{Application to a realistic minimal model}

 From the analysis of the chiral and Higgs matter contributions to the $\beta$-functions
presented above, it is clear that the cutoff independence constraints  cannot be satisfied for
any spectrum, even if  flux and other consistency conditions are met.
In this section we are going to discuss  a realistic model with the minimal spectrum
where these requirements are fulfilled.  We  make an explicit calculation to determine
 the gauge unification and $b,t$-Yukawa couplings through the renormalization
 group running  in a model which fulfills all the requirements.

To start with,  we recall here the basic features of the model~\cite{Leontaris:2010zd} which emerges from the general class~\cite{Dudas:2010zb}
presented in  Table \ref{Reps}.  The first two columns give the field content under $SU(5)\times U(1)_{t_i}$ for the case of $Z_{2}$
monodromy.  The third column presents the homology classes where $c_{1}$ is the first Chern class
of the tangent bundle of GUT surface $S_{GUT}$ and $\eta=6c_{1}-t$ with $-t$ being the first Chern class of the normal bundle
to $S_{GUT}$. The $\chi_{i}$ are unspecified subject only to the condition ${\chi}=\chi_7 +  \chi_8 + \chi_9$.  If  ${\cal F}_{Y}$ denotes
the $U(1)_Y$ flux,  to avoid a Green-Schwarz mass for the corresponding gauge boson we must require ${\cal F}_{Y}\cdot\eta={\cal F}_{Y}\cdot c_1=0$.
Then,  we get $N_i={\cal F}_{Y}\cdot\chi_i$ and  consequently $N={\cal F}_{Y}\cdot \chi= N_7+N_8+N_9$.  Using these facts, all remaining
entries of column 4 in Table \ref{Reps} are easily deduced.

We now take the flux parameters to be  $M_{10_{1,2,3}}=1$, $M_{5_{1,2,4}}=-1$
 and $N=0$, while we have the freedom to choose $N_{7,8,9}$ subject only to the constraint $N=N_7+N_8+N_9$.
This choice of $M_i,N_j$'s ensures the existence of three $10$ and three $\bar 5$ representations which are needed to
accommodate the three chiral families.
\begin{table}[tbp] \centering%
\begin{tabular}{|l|c|c|c|c|}
\hline
Field&$U(1)_i$& homology& $U(1)_Y$-flux&$U(1)$-flux\\
\hline
$10^{(1)}=10_3$& $t_{1,2}$& $\eta-2c_1-{\chi}$&$ -N$ &$M_{10_1}$\\ \hline
$10^{(2)}=10_1$& $t_{3}$& $-c_1+\chi_7$&$ N_7$ &$M_{10_2}$\\ \hline
$10^{(3)}=10_2$& $t_{4}$& $-c_1+\chi_8$&$ N_8$ &$M_{10_3}$\\ \hline
$10^{(4)}=10_2'$& $t_{5}$& $-c_1+\chi_9$&$ N_9$ &$M_{10_4}$\\ \hline
$5^{(0)}=5_{h_u}$& $-t_{1}-t_2$& $-c_1+{\chi}$&$ N$ &$M_{5_{h_u}}$\\ \hline
$5^{(1)}=5_2$& $-t_{1,2}-t_3$& $\eta -2c_1-{\chi}$&$ -N$ &$M_{5_1}$\\ \hline
$5^{(2)}=5_3$& $-t_{1,2}-t_4$& $\eta -2c_1-{\chi}$&$ -N$ &$M_{5_2}$\\ \hline
$5^{(3)}=5_x$& $-t_{1,2}-t_5$& $\eta -2c_1-{\chi}$&$ -N$ &$M_{5_3}$\\ \hline
$5^{(4)}=5_1$& $-t_{3}-t_4$& $-c_1+{\chi}-\chi_9$&$N-N_9$ &$M_{5_4}$\\ \hline
$5^{(5)}=5_{h_d}$& $-t_{3}-t_5$& $-c_1+{\chi}-\chi_8$&$ N-N_8$ &$M_{5_{h_d}}$\\ \hline
$5^{(6)}=5_y$& $-t_{4}-t_5$& $-c_1+{\chi}-\chi_7$&$ N-N_7$ &$M_{5_6}$\\ \hline
\end{tabular}%
\caption{Field representation content under $SU(5)\times U(1)_{t_i}$, their homology
class and flux restrictions~\cite{Dudas:2010zb} for the model~\cite{Leontaris:2010zd}.
Superscripts in the first column are numbering the curves, while subscripts indicate the
family, the Higgs etc. For convenience, only the properties of $10,5$ are shown.
$\ov{10},\ov{5}$ are characterized by opposite values of $t_i\ra -t_i$ etc.
Note that the fluxes satisfy $N=N_7 +  N_8 + N_9$ and $\sum_iM_{10_i}+\sum_jM_{5_j}=0$
while  ${\chi}=\chi_7 +  \chi_8 + \chi_9$.}
\label{Reps}
\end{table}

Next we use the $U(1)_Y$ flux mechanism to realise the doublet triplet splitting and
 make the model free from dangerous color triplets at
scales below $M_{GUT}$. We choose $M_{5_{h_u}}=1$, to accommodate the Higgs  ${5}_{h_{u}}$ and  $M_{5_{h_d}}=0$
with $N_8=1$  so that we get the splitting of down quark in the Higgs fiveplet
\ba
\Sigma_{5_{h_d}}:\left\{\begin{array}{ll} n_{(3,1)_{ - 1/3} }  - n_{(\overline 3 ,1)_{1/3} }& = M_{5_5 }  = 0
\\
n_{(1,2)_{1/2} }  - n_{(1,2)_{ - 1/2} } & = M_{5_5 }  + N - N_8  =  - 1
\end{array}\right.
\ea
In order to  satisfy the trace conditions we  choose $M_{5_6 }  =  - 1,\;N_7  =  - 1$
so that $\bar{5}^{(6)}$ has only a colour triplet component:
\ba
\Sigma_{5^{(6)}}:\left\{\begin{array}{ll}n_{(3,1)_{ - 1/3} }  - n_{(\overline 3 ,1)_{1/3} } & = M_{5_6 }  =  - 1
\\
n_{(1,2)_{1/2} }  - n_{(1,2)_{ - 1/2} } & = M_{5_6 }  + N - N_7  =  0\end{array}\right.
\ea
In this simple example we have succeeded to disentangle the triplet from the up-Higgs
curve at the price of generating however a new one in a different matter curve. Yet, this
allows the possibility of realising the doublet-triplet splitting since we can
generate a mass $M_{T}$ for the triplet  by coupling it to an  antitriplet via the appropriate
superpotential term~\cite{Leontaris:2010zd}. This way we obtain the corresponding Higgs doublets light.

However from Table \ref{Reps} one may see that the matter on the $\Sigma_{10^{(2,3)}}$ curves will be
affected by the $N_{7,8}$ flux. In particular the content of $10/{\ov 10}$-representations on
$\Sigma_{10^{(2,3)}}$  is split by the choice of fluxes as follows
\be
\Sigma_{10^{(2)}}:\left\{\begin{array}{ll}
n_{(3,2)_{1/6} }  - n_{(\overline 3 ,2)_{ - 1/6} }  &= M_{10_2 }  = 1
\\
n_{(\overline 3 ,1)_{ - 2/3} }  - n_{(3,1)_{2/3} }  &= M_{10_2 }  - N_7  = 2
\\
n_{(1,1)_1 }  - n_{(1,1)_{ - 1} }  &= M_{10_2 }  + N_7  = 0
\end{array}\right.\ee
\be
\Sigma_{10^{(3)}}:\left\{\begin{array}{ll}
n_{(3,2)_{1/6} }  - n_{(\overline 3 ,2)_{ - 1/6} } & = M_{10_3 }  = 1
\\
n_{(\overline 3 ,1)_{ - 2/3} }  - n_{(3,1)_{2/3} } & = M_{10_3 }  - N_8  = 0
\\
n_{(1,1)_1 }  - n_{(1,1)_{ - 1} } & = M_{10_3 }  + N_8  = 2.
\end{array}\right.
\ee
We observe that in the presence of flux one $e^c=(1,1)_1$ state is `displaced'  from $\Sigma_{10^{(2)}}$ to the
$\Sigma_{10^{(3)}}$ curve.  A similar dislocation  occurs for one $u^c=(\bar 3,1)_{-2/3}$ of $\Sigma_{10^{(3)}}$  which
`reappears' in $\Sigma_{10^{(2)}}$.  We note that this fact implies a different  structure for the up, down and charged lepton mass matrices.
It can be checked that the particular distribution of the chiral matter on the
specific  matter curves can lead to interesting results with respect to the
fermion mass structure and other phenomenological properties of the model~\cite{Leontaris:2010zd}.
For clarity, the final distribution of the MSSM spectrum along the available matter curves is summarized  in Table~\ref{content}.

Here,  we are interested for the renormalization group properties
of the model.  Using the above context we get the following relations
\begin{equation}
\label{model1}
\begin{split}
& \sum_i|M_5^i|=5,\quad \sum_i|M_5^i+N_{5Y}^i|=5\\
&\sum_j|M_{10}^j|=3,\quad \sum_j|M_{10}^j+N_{10Y}^j|=3,\quad \sum_j|M_{10}^j-N_{10Y}^j|=3
\end{split}
\end{equation}
which obviously respect the three constraints in (\ref{constraint})  and
at the same time
\be - \sum_i M_5^i= \sum_j M_{10}^j=3,\ee
which ensures three chiral families and the tracelessness condition (\ref{trace0}).

\begin{table}[!h]
\centering
\begin{tabular}{|l|r|r|rrr||l|r|r|rr|}
\hline
     &  $M$  &  $N$  &  $Q$  &  $u^c$  &  $e^c$ &     &  $M$  &  $N$  &  $d^c/T$&  $L/h_{u,d}$    \\
\hline
$10^{(1)}\,(F_3)$ &  1  &  0  &  1  &  1  &  1        &$5^{(0)}\,(h_u,T)$&  1    &  0    &  1    &  1 \\
$10^{(2)}\,(F_{2,1})$ &  1  &  -1 &  1  &  2  &  0  &$5^{(1)}\,({\bar f}_2)$  &  -1 &  0  & -1  & -1   \\
$10^{(3)}\,(F_{1,2})$ &  1  &  1  &  1  &  0  &  2  &$5^{(2)}\,({\bar f}_3)$  &  -1 &  0  & -1  & -1 \\
$10^{(4)}\,(-)$ &  0  &  0  &  0  &  0  &  0         &$5^{(3)}\,(-)$  & 0   &  0  &  0  &  0   \\
\cline{1-6}
\multicolumn{6}{c||}{}                           &$5^{(4)}\,({\bar f}_1)$  & -1  &  0  &  -1 & -1  \\
\multicolumn{6}{c||}{}                           &$5^{(5)}\,(h_d)$  & 0   &  -1 &  0  & -1  \\
\multicolumn{6}{c||}{}                           &$5^{(6)}\,(\bar T)$  & -1  &  1  & -1  & 0\\
\cline{7-11}
\end{tabular}
\caption{The distribution of the chiral and Higgs matter content of the minimal
model along the available curves, after the $U(1)_Y$ flux is turned on. The three families
  $F_{i}=10_i,\bar{f}_j=\bar 5_j$ are assigned on the curves as  indicated.
The Higgs doublets $h_{u,d}$ and   $T/\bar T$  triplets descend from three different curves.
(see also Table~\ref{Reps} and text.)}
\label{content}
\end{table}

\section{The GUT scale and the flux thresholds}

In the previous section we have shown that the contribution to the $\beta$-functions $b_a^x$ of the chiral spectrum
of the particular  $SU(5)$ model is in the required ratio, i.e. proportional to the coefficients $k_a$.

We have already explained that in the present model we assume that the breaking of $SU(5)$
occurs when a flux is turned on along the $U(1)_Y$ component of the $SU(5)$ gauge symmetry.
The rather interesting feature of the particular choice of fluxes leads to a minimal spectrum which
consists of the MSSM content and only a color triplet pair below $M_{GUT}$. Furthermore, the appearance
solely of (any number of) pairs of extra triplets -and no other additional matter- has the interesting
property of not disturbing  the scale where the unification is achieved.
This fact  is valid irrespectively of the scale that these triplets become massive.
As a result, the nice properties of the minimal supersymmetric unification scenarios where the
GUT scale is determined around $M_{GUT}\sim 10^{16}$GeV is retained.  On the other hand, the
value of the gauge coupling $g_{GUT}$ {\it does} depend on the number of triplets and the scale
they become massive.

 For convenience we start by recapitulating the basic analysis of the gauge coupling running in
the presence of fluxes. In $F$-theory constructions the flux mechanism employed to
break the GUT   symmetry  induces a splitting of the gauge couplings at the GUT
scale~\cite{Donagi:2008kj,Blumenhagen:2008aw}. The following relation is found at the unification
scale~\cite{Blumenhagen:2008aw}
\be
\frac{1}{a_Y(M_{GUT})}=\frac 53 \,\frac{1}{a_1(M_{GUT})}=\frac{1}{a_2(M_{GUT})}+\frac 23 \frac{1}{a_3(M_{GUT})}\label{SR}
\ee
The running of the couplings is governed by the RGEs where contributions to the $\beta$-functions come from the
MSSM spectrum and other possible remnants from the
higher theory. In the case of the model under consideration  we need to consider
the simple case where in addition to low energy states only  an additional triplet pair
appears in the spectrum below the $SU(5)$-GUT breaking  scale $M_{GUT}$. We
assume that at some scale $M_X<M_{GUT}$  the extra triplet pair $T,\bar T$  decouples
and only the MSSM spectrum remains massless for scales $\mu<M_X$.
The low energy values of the  gauge couplings are then given by the evolution equations
\be\label{Brun}
\frac{1}{a_a(M_Z)} = \frac{1}{a_{a}(M_{GUT})}+\frac{b_a^x}{2\pi}\,\ln\frac{M_{GUT}}{M_X}+
\frac{b_a}{2\pi}\,\ln\frac{M_X}{M_Z}
\ee
where $b_a^x, (b_a)$ are the $\beta$-functions above (below) the scale $M_X$.

Using the GUT relation (\ref{SR}) one arrives at the following formula
for the  GUT scale
\be
M_{GUT} =  e^{\frac{2\pi}{\beta {\cal A}}\rho}\,\left(\frac{M_X}{M_Z}\right)^{1-\rho}
M_Z\label{M_U}
\ee
where ${\cal A}$ is a function of the experimentally known low energy values of the
SM gauge coupling constants
\be
%\label{MGNew}
\frac{1}{\cal A} = \frac 53 \,\frac{1}{a_1(M_Z)}-\frac{1}{a_2(M_Z)}-\frac 23 \frac{1}{a_3(M_Z)}
\ee
and $\rho$ is the ratio
\be
\rho  = \frac{\beta}{\beta_x}
\ee
where $\beta,\beta_x$ are the $\beta$-functions combinations above and below the $M_X$
scale
\begin{align}
\beta_x&=b_Y^x-b_2^x-\frac 23b_3^x\label{betax}\\
\beta&=b_Y-b_2-\frac 23b_3\label{beta0}
\end{align}
If the only additional states above $M_X$ are the color triplets $T=(3,1)_{1/3},\bar T=(\bar 3,1)_{-1/3}$,
it can be easily checked that their total contribution to  $\beta_x$ combination adds up to zero. This means
that $\beta_x=\beta$ and $\rho=1$, thus  $M_{GUT}$ in (\ref{M_U}) becomes independent of
the $M_X$ scale and in fact it is identified with the MSSM unification scale
\[
M_U=M_{GUT}\equiv e^{\frac{2\pi}{\beta {\cal A}}}\,M_Z\approx 2.15\times 10^{16}\textrm{GeV}
\]
This result is only slightly modified when two loop corrections are taken into account~\cite{Leontaris:2009wi}.
The irrelevance of the triplets' decoupling scale $M_X$ holds for the one loop calculation and it is
adequate for our purposes.
With these preliminaries we are ready to discuss the computation of the Yukawa couplings.

\section{Wavefunction overlapping integrals}

In the class of models under consideration $U(1)$ symmetries are acting as family symmetries
 and  fermion families are distributed over the curves in such a way so that fermion mass textures
  are rank one. Then the only tree level Yukawa couplings are those of the third generation
  and these are computed in terms of integrals over the overlapping wavefunctions
  at the triple intersection points
  \be
\label{yukawa2}
\lambda_{ij}=M^4_*\int_S\psi_i\psi_j\phi \,\,dz_1\wedge\,d\bar z_1\wedge dz_2\wedge\,d\bar z_2
\ee
Chiral matter is localized along the matter curves and therefore the value of the integral
involving the trilinear  (top and bottom) Yukawa couplings principally depends on the
local shape of the corresponding wavefunctions.  In addition, a crucial factor in the
determination of the exact values  is  the
global normalization of the wavefunctions which could in principle depend  on the
geometry of the compact manifold. In what follows, we will make use of
the analysis of the previous section to argue that this dependence enters only
through the torsion which according to Ray-Singer theorem is independent of the
metric of the manifold.

To compute the integral we use the knowledge of the wavefunctions close to the intersection point.
The solution of the zero-mode equations leads to a Gaussian profile of the wavefunctions
$\psi$ which for localized solutions acquire  the general form
\begin{equation}
\psi
      \propto e^{-m^2\frac{|q_1z_1+q_2z_2|^2}{q}}\label{psi}
\end{equation}
where  $m$ is a mass parameter related to some background Higgs vev and $q=\sqrt{q_1^2+q_2^2}$.
We also need the normalization of the wavefunction which is determined by the integral:
\begin{equation}
{\cal C}=M_*^4\int_S\,|\psi|^2dz\wedge d\bar z\, =\,\pi \frac{M_*^4}{m^2q}{\cal R}^2\label{Cnorm}
\end{equation}
 The factor $\pi \frac{M_*^4}{m^2q}$ is the result of the gaussian integration along the coordinate normal to
the curve. The factor ${\cal R}^2$ is introduced to account for the integration along the coordinate
parametrising the curve.
We observe that in principle three different scales are introduced in the above normalization formula of the
wavefunction, namely $M_*,m$ and $ {\cal R}^{-1}$. Thus, although the solution of the equations of motion lead
to a wavefunction  which is peaked on the curve while falling off exponentially away from it, the final profile
appears to comprise  global information through the aforementioned mass parameters. We argue here that these
apparently unrelated mass scales are connected by our previous analysis and the final wavefunction formula is
solely characterised from   an invariant quantity related to the torsion.
Indeed, the scale $m$  that  has been introduced in relation to the vacuum expectation value of a background
scalar field in connection with the breaking of the enhanced gauge symmetry, is  associated to the scale   $M_*\approx M_C$
which appeared in (\ref{MGUT1}).
On the other hand, the parameter ${\cal R}$ `measures' the integration along the  matter
curve  inside  the GUT surface $S$, thus clearly ${\cal R}^{-1}<M_C$  and from our previous
analysis one naturally  expects that ${\cal R}^{-1}\approx M_{KK}\approx M_{GUT}$.
Although both scales are not known exactly, it is possible from our previous analysis
on threshold corrections to obtain a scale independent normalization of the wavefunction.
Indeed, following the reasoning presented in the above lines we can write (\ref{Cnorm}) as follows
\begin{equation}
{\cal C}=\frac{\pi}{q}\, \frac{M_C^2}{M_{GUT}^2}\label{Cnorm1}
\end{equation}
 From the $M_{GUT}$ definition formula (\ref{MGUT}) we observe that the ratio of the two scales
$M_{GUT}/M_C$  is independent of the geometry of the manifold and can be written as a simple exponential
function of the torsion. Substituting (\ref{MGUT1}) into  (\ref{Cnorm1}) we find that the wavefunctions
must be normalised with
\[
\frac{1}{\sqrt{C}}=\sqrt{\frac{q}{\pi}}\;e^{2/3({\cal T}_{5/6}-{\cal T}_0)}
\]
Therefore, the normalization constant $C$ is found to be independent of the two scales
 $M_{GUT},M_C$ of the theory.

Now, the computation of the integral for three generic  wavefunctions  of the form (\ref{yukawa2})
with arbitrary charges $q_i,q_i',q''_i, i=1,2$ participating in the triple intersection the
Yukawa couplings, gives
\begin{equation}
\lambda=e^{2({\cal T}_{5/6}-{\cal T}_0)}\frac{4\,\sqrt{\pi} }{q+q'+q''}\,\frac{(qq'q'')^{3/2}}{(q_1q_2'-q_1'q_2)^2}
\label{Yukint}
\end{equation}
where $q_i''=-q_i-q'_i$ from charge conservation in the triple intersection, while $q=\sqrt{q_1^2+q_2^2}$
and similarly for $q',q''$~\cite{Leontaris:2010zd}.
 We point out that the dependence on the value of gauge coupling $a_{GUT}$
in previous wavefunction normalizations  has been replaced by the torsion. Previous normalizations
where plagued by the smallness of the gauge coupling constant $a_{GUT}\sim\frac{1}{25}$ which led
to rather small values of the top Yukawa coupling in disagreement with the data~\footnote{
In previous normalizations, the appearance of the gauge coupling in the wavefunction normalization is due to the relation $a_{GUT}^{-1}=M_*^4\int\omega\wedge\omega \propto {\cal R}^4M_{*}^4 $. However, in this approach, the result depends on the scaling of the metric and it is not clear  whether the effective value $g_{GUT}$ or the coupling $g_s$ should be involved.}.
The hope is that the replacement of this small factor with the exponential factor incorporating the analytic
torsion might point to the correct answer.

In order to demonstrate the points discussed above, we proceed with
the computation of the determinants specifying the value of the torsion in a simple case.
As in ref~\cite{Donagi:2008kj} we take a line bundle ${\cal O}(n,-n)$ on a Hirzebruch surface
$F_0=P^1\times P^1$.  Taking into account that the Euler character on $P^{1}$ is given
${\chi}(P^1,{\cal O}(n))=n+1$ the product formula of Ray-Singer torsion yields
\[{\cal T}_{{\cal O}(n,m)}=(n+1){\cal T}_{{\cal O}(m)}+(m+1){\cal T}_{{\cal O}(n)}\]
Since we want to eliminate the color triplets $(3,2)_{-5/6}+(\bar 3,2)_{5/6}$ we need to
take $\chi(P^1,{\cal O}(n))=n+1=0$, therefore for the line bundle ${\cal L}^{5/6}$ we take $n=-1$,
so
\[{\cal T}_{5/6}={\cal T}_{{\cal O}(1,-1)}=2{\cal T}_{{\cal O}(-1)}+(-1+1){\cal T}_{{\cal O}(1)}=2{\cal T}_{{\cal O}(-1)}\]
For the trivial line bundle we take $n=0$, hence
\[{\cal T}_{0}={\cal T}_{{\cal O}(0,0)}=2{\cal T}_{{\cal O}(0)}\]

We now recall that a positive  elliptic differential operator, as is the case of
$\Delta_{k,R}$ with spectrum $\lambda_n$, is associated to the zeta functions~\cite{Ray:1971,Nash:1992sf}
\[\zeta_{k}(s)=\sum_{\lambda_n}\frac{\gamma_n}{\lambda_n^s}\]
where $\gamma_n$ is the degeneracy of $\lambda_n$. The determinant is defined by
\[\log({\rm Det}\Delta)=-\left.\frac{d\zeta_{k}(s)}{ds}\right|_{s=0}\]
Therefore,  the torsion is given ${\cal T}_{{\cal O}(k)}=-\frac 12\zeta'_k(0)$ with
\[\zeta'_k(0)=4\zeta_R'(-1)-\frac 12(k+1)+\sum_{l=1}^{k+1}(2l-|k+1|)\log l\]
where $\zeta_R'(-1)\approx -0.165421$.
Then we get
\[{\cal T}_{5/6}-{\cal T}_0=2({\cal T}_{{\cal O}(-1)}-{\cal T}_{{\cal O}(0)})=\zeta'_0(0)-\zeta'_{-1}(0)=-\frac 12\]
For this particular choice of bundle, the Yukawa coupling then becomes
\begin{equation}
\lambda=\frac{4\sqrt{\pi}/e }{q+q'+q''}\,\frac{(qq'q'')^{3/2}}{(q_1q_2'-q_1'q_2)^2}
\label{Lambda}
\end{equation}

\subsection{The top mass}

Let us now apply the above formula in the case of the model presented in the previous section.
 From  Table  \ref{Reps}  we see that a tree-level mass for the top quark is available
from the gauge invariant Yukawa coupling
\begin{align}
W_{tree}&=\lambda_t\;{\bf 10}_3\;\;\cdot\; {\bf 10}_{3}\;\;\;\cdot\;\;\; {\bf  5}_{h_u}\nn\\
&          \quad\quad\;\; \; \;t_1\;\;\;\;\;\;\;\; t_2\;\;\;\;-t_1-t_2\nn
\end{align}
where from the second line we see that the $U(1)$ invariance  $\sum_it_i=0$ is also satisfied.
The charges in (\ref{Lambda}) for the top vertex for this model have been calculated in \cite{Leontaris:2010zd}
\ba
\{q_1,q_2\}=\left\{\sqrt{\frac{3}{10}},\frac{1}{\sqrt{2}}\right\},
\;\{q_1',q_2'\}=\left\{\sqrt{\frac{3}{10}},-\frac{1}{\sqrt{2}}\right\}
\label{Qtop}
\ea
The resulting top quark coupling is computed form (\ref{Lambda}) and is found
 to be $\lambda_t\approx 1.23.$
Similarly  the bottom quark Yukawa coupling in the model is  obtained  form the superpotential term
\begin{align}
W_{tree}&=\lambda_b\;{\bf 10}_3\;\;\cdot\;\; {\bf \bar 5}_{3}\;\;\;\cdot\;\;\; {\bf \bar 5}_{h_d}\nn\\
&            \quad\quad\;\;\;\;t_2\;\;\;\; t_3+t_5\;\;t_1+t_4\nn
\end{align}
while repeating the above analysis one finds $\lambda_b\approx 1.17$. We thus observe
that  we are in the large $\tan\beta$ regime with a top coupling in the range $\lambda_t
\ge 1$ which implies a  physical mass around $186$ GeV
which is within the experimental  range~\cite{Aaltonen:2011wt}.

\begin{figure}[!t]
\centering
\includegraphics[scale=0.68]{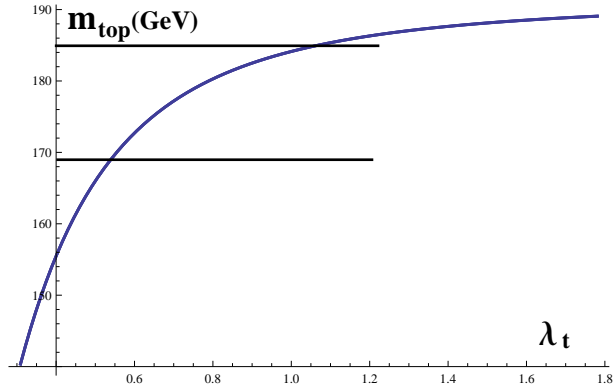}
\caption{Plot of the top mass versus the value of the Yukawa coupling $\lambda_t(M_{GUT})$,
for $\tan\beta=(40-55)$. Horizontal lines indicate the experimental uncertainties of the top mass. }
\label{lambdat}
\end{figure}

\begin{figure}[!b]
\centering
\includegraphics[scale=0.68]{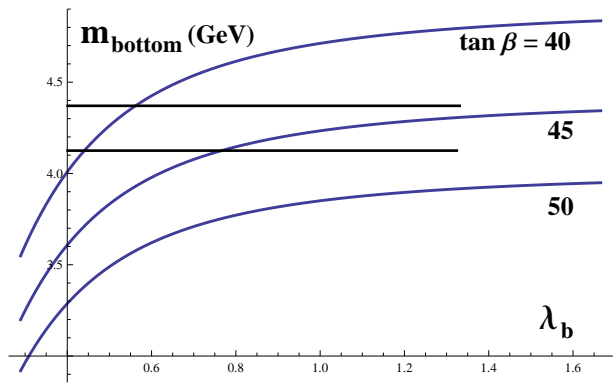}
\caption{Plot of the bottom mass versus the value of the Yukawa coupling $\lambda_b(M_{GUT})$
for three different values of  $\tan\beta=(40,45,50)$. Horizontal lines indicate the experimental uncertainties on the bottom mass. }
\label{lambdab}
\end{figure}

In deriving the above numerical  values we have made a specific simple choice of flat line bundle to compute the
torsion  which is involved in the top and bottom mass formula  through the exponential factor  appearing in (\ref{Yukint}).  Of course, one
may think of more involved cases of bundle structures however the above estimates are not expected to
alter dramatically.  Moreover, in a fully realistic model  other parameters emerging from various thresholds
in the renormalization group running etc may modify also the computation in obtaining the low energy masses.
The final result will also depend on the specific choice of $\tan\beta$  which is  fixed only by phenomenology and {\it not} by some theoretical principle.  Thus,
in our present computations we have the freedom to adjust its value to fit the low energy mass data.
Furthermore,  the reliability of our computations is corroborated  from
the  rather interesting property of the top coupling, namely  its fixed point behavior.
This means that for a range of relatively large initial $\lambda_t$ values at GUT, the renormalized
value of the top Yukawa approaches the same limiting value at the scale of $M_W$. It is anticipated
that another choice of bundle   would imply reasonable deviations from the numbers obtained in this particular example, so that they do not alter substantially the low energy predictions.
 In this sense, a fully realistic model could arise under an appropriate type  of bundle structure.
For completeness, we demonstrate the fixed point behavior  by a one-loop renormalization
group analysis using analytic formulae~\cite{Floratos:1995dh} of the top-bottom  Yukawa coupled differential equations in
the Appendix. To visualize this effect, using the derived formulae,
in Fig.(\ref{lambdat}) and (\ref{lambdab}) we plot the low energy top and bottom masses as a function of the values of the
corresponding Yukawa couplings
at $M_{GUT}$.  This is done for a range of large values of $\tan\beta\approx (40-55)$ as indicated in the plots.
We can see clearly in the bottom mass plot the three curves corresponding to the values $\tan\beta =40,45,50$. Notice that in the
case of the top mass plot, all $\tan\beta$ curves are squeezed to one because the top mass depends on $\sin(\beta)\lesssim {\cal O}(1)$.
Horizontal lines specify the experimental uncertainties.
It can be seen that top mass is found in a region where the Yukawa coupling is fairly  close enough
to  the fixed point and small uncertainties in its computation are within the experimentally acceptable bounds.

\section{Conclusions}

In this work we have reconsidered the possibility of determining in a reliable way
the mass scales and moduli parameters which enter in the computation of
 the tree level Yukawa couplings in a class of F-Theory GUT models.  We consider
 in particular the $SU(5)$ GUT model generated by a seven brane wrapping a complex surface
  $S$ and embedded in the exceptional $E_8$ underlying theory. The $SU(5)$ GUT breaking
  is supposed to occur when a $U(1)_Y$ flux is turned on. In these models
 chiral fields are accommodated along the Riemann surfaces which constitute the
 intersections of the GUT brane with other 7 branes where the $SU(5)$ gauge symmetry is
 enhanced. Chiral matter fields lying on these `curves' are charged under additional
$U(1)$'s    which act as family symmetries. Under certain monodromy actions, Yukawa couplings can be formed  at triple
intersections of such curves and predict rank one fermion mass textures. The values
of the Yukawa couplings are expressed in terms of integrals of overlapping wavefunctions
of the matter fields participating in the triple intersection. The wavefunctions are found from the solution of the equations
of motion and their normalization in principle appears to require knowledge of the mass scales of the theory and topological
properties of the compact space and matter curves. We have made reasonable
assumptions about the mass scales of the theory and used previous analysis of threshold
effects from KK modes~\cite{Friedmann:2002ty,Donagi:2008kj} to  determine the
GUT scale and the mass ratios which control the wavefunction normalization.
Further, we have studied similar effects from a generic chiral sector of the F-theory model in the presence
of $U(1)$ fluxes and have analysed the constraints imposed on the light spectrum   of the effective field theory model.
 We have incorporated the above findings into the computation of  generic trilinear
 top and bottom Yukawa couplings in the context of the $SU(5)$ model embedded in $E_8$
 and found that  the couplings depend only on the extra $U(1)$ charges carried by the
 fields at the intersection and a topological invariant quantity, namely the Ray-Singer torsion~\cite{Ray:1973sb}.
  Furthermore, we have applied the method of our computation in a viable model obtained in the
 described F-theory context~\cite{Leontaris:2010zd} with the minimal light spectrum
 which meets all the requirements imposed by our analysis. To compute the torsion, we
have used~\cite{Donagi:2008kj} as example a simple case of flat line bundle on a
 Hirzebruch surface.   We have performed  the renormalization group  analysis, taking into account
 the KK-mode effects discussed previously as well as
 flux thresholds to determine the low energy values of the gauge and Yukawa couplings.
 We have found a top and bottom mass which is in  agreement with their
 experimentally determined values.

 \bigskip

 \noindent{\bf{\Large Acknowledgement}}

 We would like to thank CERN theory group for kind hospitality during the last stage of this work.
This work is partially supported by the European Research and Training Network (RTN)
grant ``Unification in the LHC era ''(PITN-GA-2009-237920).

\newpage
\noindent {\bf Appendix }

In subsection 6.1 we applied our method for the computation of the top and bottom Yukawa
couplings which for a specific choice of line bundle they where  found to obtain ${\cal O}(1)$
and comparable values (large $\tan\beta$ regime) at the GUT scale.
 The top and bottom masses are given  from their renormalised values at scales
$\sim M_W$, and are obtained from a solution of two coupled differential equations (DEs)
(we assume for  simplicity small tau Yukawa coupling).
 For comparable initial $\lambda_{t,b}$ values the DEs cannot be disentangled, thus in order to give a
 reliable calculation of their low energy values, we  need to solve the above DEs and plot the values
 of the top and bottom masses for reasonable ranges of $\lambda_{b,t}$ initial values.
To this end, in this Appendix we obtain analytic formulae by
extending the analytic solution~\cite{Floratos:1994ap}  of the system of the 1-loop RGE
for the top and bottom Yukawa couplings $h_t$ and $h_b$. Defining
$\ga_t=\lambda_t^2/(4\pi)$ and $\ga_b=\lambda_b^2/(4\pi)$, the RGEs are
\begin{equation}
\label{dif_eq}
\begin{split}
\frac{d\ga_t}{dt}&=\frac{1}{2\pi}\left(6\ga_t+\ga_b-G_Q\right)\ga_t\\
\frac{d\ga_b}{dt}&=\frac{1}{2\pi}\left(\ga_t+6\ga_b-G_D\right)\ga_b
\end{split}
\end{equation}
with $G_I=c^I_1 \ga_1+c^I_2 \ga_2+c^I_3 \ga_3$, $I=Q,D$, and the constants $c$ are given by
($SU(5)$ normalization)
\[
\{c^Q_1,c^Q_2,c^Q_3\}=\left\{\frac{16}{3},3,\frac{13}{15}\right\},\quad\quad\textrm{and}\quad
\{c^D_1,c^D_2,c^D_3\}=\left\{\frac{16}{3},3,\frac{7}{15}\right\}
\]
Ignoring the small difference in the $U(1)$ factor of the constants $c$, we shall make the transformations
\begin{equation}
\label{trans_1}
\ga_t=\gamma^2_Q x,\quad\quad\textrm{and}\quad \ga_b=\gamma^2_D y\sim \gamma^2_Q y
\end{equation}
where
\[
\gamma^2_Q=\exp\left[-\frac{1}{2\pi}\int_{t_0}^t G_Q(t')dt'\right]
\]
The system (\ref{dif_eq}) becomes
\begin{equation}
\label{dif_eq_2}
\begin{split}
\frac{dx}{dt}&=\frac{1}{2\pi}\left(6x+y\right)\gamma^2_Q x\\
\frac{dy}{dt}&=\frac{1}{2\pi}\left(x+6y\right)\gamma^2_Q y
\end{split}
\end{equation}
Simple manipulations give
\begin{equation}
\label{dif_eq_3}
\begin{split}
\frac{d\ln(x-y)}{dt}&=\frac{\gamma^2_Q}{2\pi}\,6(x+y)\\
\frac{d\ln(xy)}{dt}&=\frac{\gamma^2_Q}{2\pi}\,7(x+y)
\end{split}
\end{equation}
Therefore
\begin{equation}
\label{function}
7\frac{d\ln(x-y)}{dt}=6\frac{d\ln(xy)}{dt},\quad\textrm{or}\quad
\left(\frac{xy}{x_0y_0}\right)^6=\left(\frac{x-y}{x_0-y_0}\right)^7
\end{equation}
where $x_0$ and $y_0$ are initial conditions at the scale $t_0$. Observe that, since $\gamma^2_Q(t_0)=1$,
we have that $x_0=\ga_t(t_0)$ and $y_0=\ga_b(t_0)$.
Using (\ref{function}) we can write
\[
(x+y)^2=(x-y)^2+4xy=(x-y)^2+4\frac{x_0y_0}{(x_0-y_0)^{7/6}}\,(x-y)^{7/6}
\]
and the first of the equation in (\ref{dif_eq_3}) reads
\[
\frac{d(x-y)}{dt}=\frac{\gamma^2_Q}{2\pi}\,6(x-y)\sqrt{(x-y)^2+4\frac{x_0y_0}{(x_0-y_0)^{7/6}}\,(x-y)^{7/6}}
\]
Defining a further new transformation
\begin{equation}
\label{trans_2}
u=k_0/\omega^{5/6}=k_0/(x-y)^{5/6},\quad\quad\textrm{where}\quad k_0=4\frac{x_0y_0}{(x_0-y_0)^{7/6}}
\end{equation}
we get
\[
\begin{split}
\frac{du}{dt}&=(-\frac 56)\frac{k_0}{\omega^{11/6}}\frac{d\omega}{dt}=
-\frac{5}{2\pi}\frac{k_0}{\omega^{11/6}}\gamma^2_Q\omega\sqrt{\omega^2+k_0\omega^{7/6}}\\
&=-\frac{5}{2\pi}\gamma^2_Q k_0\frac{u}{k_0}\sqrt{\left(\frac{k_0}{u}\right)^{12/5}+k_0\left(\frac{k_0}{u}\right)^{7/5}}\\
&=-\frac{5}{2\pi}\gamma^2_Q k_0^{6/5} u \frac{\sqrt{1+u}}{u^{6/5}}\\
&=-\frac{5}{2\pi}\gamma^2_Q k_0^{6/5}\frac{\sqrt{1+u}}{u^{1/5}}
\end{split}
\]
Therefore
\begin{equation}
\label{dif_eq_4}
\frac{u^{1/5}}{\sqrt{1+u}}\,du=-\frac{5}{2\pi} k_0^{6/5}\gamma^2_Q dt
\end{equation}
Integrating the left-hand side we get
\[
\int \frac{u^{1/5}}{\sqrt{1+u}}\,du=
\frac{10}{7}\,u^{1/5}\left[\sqrt{1+u}-\,_2\mathcal{F}_1\left(\frac 15,\frac 12,\frac 65,-u\right)\right]
\]
where $_2\mathcal{F}_1$ is the hypergeometric function. Let us work on the right-hand side of (\ref{dif_eq_4})
\[
\gamma^2_Q=\exp\left[-\frac{1}{2\pi}\int_{t_0}^t G_Q(t')dt'\right]=
\exp\left[-\frac{1}{2\pi}\int_{t_0}^t \left(c^Q_1 \ga_1(t')+c^Q_2 \ga_2(t')+c^Q_3 \ga_3(t')\right)dt'\right]
\]
Since
\[
\int_{t_0}^t\ga_i(t')\,dt'=\int_{t_0}^t\frac{\ga_{i0}\,dt'}{1-\frac{b_i\ga_{i0}}{2\pi}(t'-t_0)}=
-\frac{2\pi}{b_i}\ln\left[1-\frac{b_i\ga_{i0}}{2\pi}(t-t_0)\right]
\]
we get
\[
\gamma^2_Q=\prod_{i=1}^3 \left[1-\frac{b_i\ga_{i0}}{2\pi}(t-t_0)\right]^{c^Q_i/b_i}
\]
Define for convenience the new constants (we suppress hereafter the superscript $Q$)
\[
b_i\ga_{i0}/(2\pi)=B_i,\quad\quad\textrm{and}\quad c_i/b_i=e_i
\]
Integrating $\gamma^2_Q$ we get
\[
\begin{split}
&\int\gamma^2_Q dt=\frac{1}{B_1(1+e_1)}
\left[-1+B_1(t-t_0)\right]
\left[1-B_1(t-t_0)\right]^{e_1}
\left[1-B_2(t-t_0)\right]^{e_2}\times\\
&\left[\frac{B_1}{B_1-B_2}\left(1-B_2(t-t_0)\right)\right]^{-e_2}
\left[1-B_3(t-t_0)\right]^{e_3}
\left[\frac{B_1}{B_1-B_3}\left(1-B_3(t-t_0)\right)\right]^{-e_3}\times\\
&_A\mathcal{F}\left(1+e_1,-e_2,-e_3,2+e_1,\frac{B_2(-1+B_1(t-t_0))}{B_1-B_2}, \frac{B_3(-1+B_1(t-t_0))}{B_1-B_3}\right)
\end{split}
\]
where $_A\mathcal{F}$ is the Appell hypergeometric function of two variables~\cite{Appel}.
Simplifying the relation and using $1-B_1(t-t_0)=\ga_{10}/\ga_1(t)$, we get
\[
\begin{split}
&\int\gamma^2_Q dt=\frac{1}{B_1(1+e_1)}
\left[-\left(\frac{\ga_{10}}{\ga_1(t)}\right)^{e_1+1}\right]\left[\frac{B_1-B_2}{B_1}\right]^{e_2}\left[\frac{B_1-B_3}{B_1}\right]^{e_3}\times\\
&_A\mathcal{F}\left(1+e_1,-e_2,-e_3,2+e_1,
\frac{B_2}{B_2-B_1}\left(\frac{\ga_{10}}{\ga_1(t)}\right),\frac{B_3}{B_3-B_1}\left(\frac{\ga_{10}}{\ga_1(t)}\right)\right)
\end{split}
\]

\noindent
Therefore the differential equation (\ref{dif_eq_4}) gives
\[
\begin{split}
&\frac{10}{7}\,u^{1/5}\left[\sqrt{1+u}-\,_2\mathcal{F}_1\left(\frac 15,\frac 12,\frac 65,-u\right)\right]=\\
&-\frac{5}{2\pi}k_0^{6/5}\frac{1}{B_1(1+e_1)}
\left[-\left(\frac{\ga_{10}}{\ga_1(t)}\right)^{e_1+1}\right]\left[\frac{B_1-B_2}{B_1}\right]^{e_2}\left[\frac{B_1-B_3}{B_1}\right]^{e_3}\times\\
&_A\mathcal{F}\left(1+e_1,-e_2,-e_3,2+e_1,
\frac{B_2}{B_2-B_1}\left(\frac{\ga_{10}}{\ga_1(t)}\right),\frac{B_3}{B_3-B_1}\left(\frac{\ga_{10}}{\ga_1(t)}\right)\right)+\textrm{constant}
\end{split}
\]
and simplifying
\begin{equation}
\label{gen_sol}
\begin{split}
&u^{1/5}\left[\sqrt{1+u}-\,_2\mathcal{F}_1\left(\frac 15,\frac 12,\frac 65,-u\right)\right]=\\
&\frac{7}{4\pi}
k_0^{6/5}\frac{1}{B_1(1+e_1)}
\left[\left(\frac{\ga_{10}}{\ga_1(t)}\right)^{e_1+1}\right]\left[\frac{B_1-B_2}{B_1}\right]^{e_2}\left[\frac{B_1-B_3}{B_1}\right]^{e_3}\times\\
&_A\mathcal{F}\left(1+e_1,-e_2,-e_3,2+e_1,
\frac{B_2}{B_2-B_1}\left(\frac{\ga_{10}}{\ga_1(t)}\right),\frac{B_3}{B_3-B_1}\left(\frac{\ga_{10}}{\ga_1(t)}\right)\right)+\textrm{constant}
\end{split}
\end{equation}
Let us see now how $x$ and $y$ are related with $\omega$ which, from (\ref{trans_2}) is given in terms of $u$. Since
(see (\ref{function}))
\[
x-y=x+(-y)=\omega,\quad\quad\textrm{and}\quad (-y)\cdot x=-x_0y_0\left(\frac{\omega}{\omega_0}\right)^{7/6}
\]
we get (using (\ref{trans_2}) for the second equality)
\begin{equation}
\label{x_y}
\begin{split}
x&=\frac{\omega}{2}\left(1+\sqrt{1+k_0\omega^{-5/6}}\right)=\frac 12 \left(\frac{k_0}{u}\right)^{6/5}\left(1+\sqrt{1+u}\right)\\
y&=\frac{\omega}{2}\left(-1+\sqrt{1+k_0\omega^{-5/6}}\right)=\frac 12 \left(\frac{k_0}{u}\right)^{6/5}\left(-1+\sqrt{1+u}\right)
\end{split}
\end{equation}
 From (\ref{trans_1}) we get
\begin{equation}
\begin{split}
\ga_t&=\gamma^2_Q\,x=\prod_{i=1}^3 \left[1-\frac{b_i\ga_{i0}}{2\pi}(t-t_0)\right]^{c_i^Q/b_i}
  \frac 12 \left(\frac{k_0}{u}\right)^{6/5}\left(1+\sqrt{1+u}\right)\\
\ga_b&=\gamma^2_D\,y=\prod_{i=1}^3 \left[1-\frac{b_i\ga_{i0}}{2\pi}(t-t_0)\right]^{c_i^D/b_i}
  \frac 12 \left(\frac{k_0}{u}\right)^{6/5}\left(-1+\sqrt{1+u}\right)
\end{split}
\end{equation}
These formulae incorporate the RG running effects and encompass all the information for the initial
values  of the gauge and Yukawa couplings at the GUT scale. We may substitute the
top and bottom initial values and use them~\cite{Floratos:1995dh} to derive our results.

\end{document}